\documentclass[twocolumn,aps,prb,superscriptaddress,amsmath,amssymb,
showpacs,eqsecnum,floatfix]{revtex4}

\usepackage{graphicx}
\usepackage{dcolumn}
\usepackage{bm}

\begin{document}


\title{Thermodynamics and Phase Diagrams  of layered superconductor/ferromagnet
nanostructures}

\author{Paul H. Barsic}
 \email{barsic@physics.umn.edu}

\author{Oriol T. Valls}
\email{otvalls@umn.edu}
\affiliation{ School of Physics and Astronomy, University of Minnesota, Minneapolis, Minnesota 55455}
 \altaffiliation[Also at ]{Minnesota Supercomputer Institute, University of Minnesota, Minneapolis, Minnesota 55455}

\author{Klaus Halterman}
\email{klaus.halterman@navy.mil}
\affiliation{Physics and Computational Sciences, Research and Engineering Sciences Department, Naval Air Warfare Center, China Lake, California 93555, USA}

\date{\today}

\begin{abstract}
We study the thermodynamics of 
clean, layered superconductor/ferromagnet nanostructures using fully self
consistent methods to solve the microscopic Bogoliubov-deGennes
equations. From these self-consistent
solutions the condensation free energies are obtained.
The trilayer SFS junction is studied in particular detail: 
first 
order transitions between $0$ and $\pi$ states as
a function of the temperature $T$ are located by finding
where the free energies of the two phases cross. The occurrence  of these
transitions
is mapped as a function
of the thickness $d_F$ of the F layer and of the Fermi wavevector mismatch
parameter $\Lambda$.  Similar first order transitions are found for systems
with a larger number of layers: examples are given in the 7 layer (3
junction) case.
The latent heats associated with these phase transitions are evaluated and 
found to be experimentally accessible. 
The transition temperature to the normal
state is calculated from the linearized Bogoliubov-deGennes equations and 
found to be in good agreement with experiment.
Thus, the whole three dimensional phase diagram in $T,d_F,\Lambda$ space
can be found. The first order transitions are associated with dips in
the transition temperature $T_c$ to the non-superconducting state, which should
facilitate locating them.
Results are given also for the magnetic moment and the local
density of states (DOS) at the first order transition.
\end{abstract}

\maketitle

\section{\label{sec:intro}Introduction}
The investigation of systems
involving ferromagnet (F) and superconductor (S) junctions
is an active component of superconductor-based 
spintronics\cite{igor} research.  
A broad array of interesting effects arises in S/F nanostructures,
which
opens  doors for nanotechnologies and  associated
devices and applications that may offer benefits beyond current superconducting
devices such as standard Josephson junctions.
Advances in fabrication techniques permit growth of
ferromagnet and superconductor layers in the form of junctions and 
heterostructures smooth up to the atomic scale.

The arrangement of consecutive F and S layers, as in SFS 
junctions, results in competition between magnetic and superconducting
orderings.  Superconducting correlations can leak
into the ferromagnet while spin polarization can extend into the superconductor:
these are the now well established S/F proximity 
effects.\cite{andreev,buzdin_rmp}
The phase coherence embodied in the superconducting correlations
becomes modified in the F regions. 
The exchange energy in the ferromagnet shifts the kinetic energies of the 
quasiparticles constituting the Cooper pairs and subsequently a new 
superconducting state arises whereby the center of mass momentum of the pair 
is nonzero.\cite{demler}
This results in 
a spatially decaying  pair amplitude that oscillates over a 
characteristic length scale much smaller than the superconducting coherence 
length.  
The modulating pair amplitude within the magnet indirectly links adjacent
S layers, and thus proximity effects in F cause local oscillations 
in  physically relevant single-particle quantities, including the
magnetization\cite{koshina,hv04} and density of states\cite{hv02,buzdin_dos} 
(DOS).  Similarly, in the S material the magnet locally  polarizes the 
superconductor, causing a monotonic decline in the pairing correlations 
near the interface over an extended region. 
The associated spin-split Andreev quasiparticle states also lead to
interesting local behavior in the DOS and magnetic moment in the superconductor.
The nontrivial behavior of the proximity effects in these 
structures  plays a central role in 
the competition between the magnetic and superconducting order.

The modification of the superconducting phase coherence due to proximity 
effects in clean multilayers consisting of one or more successive SFS 
junctions is particularly striking.  On the atomic level, the  pair amplitude 
is a smoothly varying  function of the spatial coordinates. 
Depending on the values
of certain  parameters (such as
F layer width, $d_F$) the  damped oscillatory pair amplitude in the 
F layer may arrange itself in such a manner that is energetically favorable for
its sign 
to change from one of the S layers to the next,
yielding a so-called $\pi$-junction, as first
proposed long ago.\cite{bula} If the pair amplitude 
does not change sign between S layers, it is an ordinary or 
$0$-junction.  There is a rich and broad  parameter space that then enables a 
certain level of control over the competing magnetic and superconducting 
orderings, allowing one to increase or diminish the proximity effects that 
dominate the relative SFS coupling.  The actual equilibrium state ($0$ or $\pi$)
is dependent upon several variables, including predominantly
the F region's material characteristics and the temperature, $T$, 
all of which ultimately determine the pair amplitude modulation in the magnet.
A system comprised of a larger number of SFS sequences 
results in a greater number of possible $0$ or $\pi$ junction combinations.

The transitions between $0$ and $\pi$ states can be 
explored through the signatures of a variety of physical parameters.
Experimental study of this question 
has focused primarily on measurements
of 
the critical current $I_c$\cite{ryazanov,sellier,frolov,guichard,jin,milosevic,frolov2,kontos2,oboznov}
and, thermodynamically, 
on the critical temperature\cite{pokrovsky,obi99,jiang95,jiang96,muhge96,shelukhin} 
of the transition to the normal state, $T_c$. 
Evidence of $0\leftrightarrow\pi$ transitions can be seen in the SFS Josephson 
coupling, which 
manifests itself in the vanishing of  $I_c$, although 
higher order harmonics in the current-phase relationship can modify 
this.\cite{radovic}
Measurements\cite{pokrovsky,obi99,jiang95,jiang96,muhge96,shelukhin}  
as  a function of $d_F$ have shown that 
$T_c$,
which is of course smaller than $T_c^0$, 
the critical temperature
for  bulk S material, oscillates as a function of $d_F$, 
confirming theoretical
predictions\cite{radovic0,krun} 
based upon the semi-classical 
Usadel equations.  
Intrinsically linked to this phenomenon  are  damped 
oscillations in $I_c$  as a function of $d_F$ and 
exchange energy
in the clean\cite{buzdin} and dirty limits.\cite{buzdin2,buzdin3}
These changes in the critical current have been experimentally 
confirmed\cite{ryazanov,sellier,frolov,guichard,jin,milosevic,kontos2,oboznov,frolov2}.
Of particular interest is Ref.~\onlinecite{frolov2}, which demonstrates
the robustness of $0\leftrightarrow\pi$ transitions by providing evidence of switching in
samples with interfaces that were not atomically smooth.
Indeed, despite deviations as large as $0.6 {\rm nm}$ 
over $10\%$ of a sample,  
clear evidence of switching was found. 
Near $T_c$,  and in the diffusive limit, the theory
was later extended to include arbitrary interface transparency.\cite{buzdin3}
Measurements of the superconducting phase\cite{guichard} have corroborated 
the $\pi$ state in SFS junctions, and the predicted oscillations in several 
thermodynamic quantities have in many cases been found experimentally.
Direct evidence of DOS oscillations was reported in 
a tunneling spectroscopy experiment,\cite{kontos} but not observed\cite{beasley,courtois} in other cases. 
Such studies give us the valuable insight that  
the oscillations 
are correlated with  $\pi \leftrightarrow 0$ transitions. 
In this work, we show that there is indeed an intimate 
relation between the oscillations 
in $T_c$ as a function of relevant
parameters and the transitions from the $\pi$ to the $0$
state and we find good
quantitative agreement with experimental data.

Since the possibility of having a particular junction configuration depends
fundamentally on the intricate properties of the pair amplitude, 
the complicated and demanding task of calculating 
the pair potential, $\Delta({\bf r})$, 
rigorously and self-consistently becomes 
absolutely necessary, particularly as the inhomogeneities occur
on  a microscopic scale.
The first step in the self-consistency process often involves 
an assumed simple piecewise constant form for $\Delta({\bf r})$,
which is then iterated through the relevant  equations until 
convergence is achieved.  It is not justified
to bypass the technical difficulties associated with self-consistency and
to use only an assumed form for the pair potential.
The final calculated $\Delta({\bf r})$ often deviates significantly, even
in overall symmetry, from
the assumed form: the self-consistent $\Delta({\bf r})$
has a complicated spatial behavior that can  lead to  stable 
states mixing $0$ and $\pi$ junction configurations.\cite{hv03} 
A self-consistently calculated pair potential
minimizes, at least locally, the free energy 
of the system. To determine whether
the calculated state is merely a local minimum of the free energy 
or the global one, the free energies from all 
possible self-consistent  $0$- and  $\pi$-junction
configurations must be compared with high precision.
Recently developed numerical algorithms\cite{hv03,bhv06} overcome
the difficulties that arise in computing the small difference between
much larger quantities and enable accurate computation of the   
differences in the values of the condensation free energy
of different minima.

For clean SFS junctions,
a relevant set of basic parameters to consider includes
$d_F$, the exchange energy $h_0$ 
and $T$.
As these parameters vary, 
the $0$ or $\pi$-state free energies may cross at certain 
points in parameter space, yielding phase transitions.
It has been shown\cite{hv03} that at $T=0$, 
transitions occur when varying
$h_0$, $d_F$ and also the mismatch parameter $\Lambda$, defined as the  
ratio of Fermi energies in the F and S regions. This
mismatch can induce a transition because at $\Lambda \approx 1$, when
the Fermi wavevectors match, 
the layers couple more strongly, while at small $\Lambda$ 
the coupling is effectively weaker.
If the temperature varies it is also possible 
to have a first order transition between $0$ and $\pi$ junction states, as
recently shown in both the clean,\cite{bhv06} and dirty\cite{tollis} limits,
and
also predicted for short-period F/S superlattices.\cite{radovic0}
The temperature has been shown to have a pronounced effect on
the pair amplitude in the F region of F/S structures,\cite{hv02a}
strongly diminishing its magnitude  while maintaining
its characteristic period of oscillation as $T$ increases. This 
translates into weaker coupling between adjacent S layers.
If the magnet width is such that
the junction is near  a $0\leftrightarrow\pi$ transition point at 
$T=0$, increasing the temperature can result in the critical current of the 
junction having a non-monotonic temperature dependence.\cite{radovic}
It has been argued\cite{cayssol} that the transition is discontinuous in
uniform samples but rounded off in samples of variable thickness.\cite{buzdin4}
However, a transition can be
observed\cite{frolov2} in just a portion
of samples with nonuniform thickness.
These results  indicate that the temperature 
can be used to switch between a $0$ and  a $\pi$ state configuration. 
It is possible to locate regions of parameter space that give the desired 
transitions using the $T=0$ results as guides, however the task is still 
significantly demanding.  
Such temperature transitions were found to occur in one-junction 
and 3-junction systems for moderate 
values of $\Lambda$.\cite{bhv06}
Thus, a 1-junction system was  
found to have a $0\rightarrow \pi$ first order transition 
as $T$ was lowered, and a $\pi\pi\pi \rightarrow \pi0\pi$
transition was found for a
3-junction system.\cite{bhv06} 
In each case, the free energies of a stable
and a metastable state crossed at the
transition temperature with differing derivatives, and therefore entropies.
The existence of metastable states and an entropy
discontinuity are hallmarks of first-order phase transitions.
Moreover, the 
reported latent heats were reported to be  within available
experimental resolution. It is therefore desirable to systematically 
study the coexistence of metastable
states and the nature of the transition in SFS and higher-order multilayer structures.

The main objective of this paper therefore,
is to map out the 
regions of parameter space in which the different
junction states are stable, and to trace the locations of the phase 
transitions  in systems with SFS junctions. 
An extensive sweep of the geometric and materials parameters 
including $d_F$, $\Lambda$, as well as $T$, is performed. 
To start with, it is
important to know which $d_F$ and $\Lambda$ ranges allow 
more than one self-consistent
state at $T=0$. One can
then check if a metastable state at low temperature  becomes 
the equilibrium state at higher $T$. 
By using this procedure we obtain a complete phase diagram of an SFS 
junction within the relevant region of $(T,\Lambda,d_F)$ space.
To accomplish this, we use a method that can accommodate arbitrary values 
of the above parameters, without recourse to approximations.
As discussed 
above, all calculations involving the pair potential 
must be performed using fully self-consistent algorithms, 
starting from the microscopic equations (Bogoliubov-deGennes (BdG)).
The need for a fully microscopic theory arises
because the characteristic period of the pair potential oscillations 
approaches the atomic scale.
For the nanoscale interlayer widths considered here, 
geometrical oscillations decisively influence the
final results.

We present in Sec. \ref{sec:methods} the microscopic  equations and the 
associated notation relevant for systems containing SFS junctions.
We review the numerical procedures involved in calculating the 
self-consistent pair potential and quasiparticle spectra, and
the method used to 
calculate the primary thermodynamic quantity, the condensation free energy, 
$\Delta {\cal F}(T)$, from the self-consistent spectrum and pair potential.
We also outline a semi-analytic method to calculate
$T_c$ through the linearized BdG equations.
In Sec. \ref{sec:results}, we show that first order transitions
with measurable latent heat
can occur between states containing different
numbers of $0$ and $\pi$ junctions as the temperature changes.
For SFS junctions the transitions we find
are from the $\pi$ to the $0$ state as $T$ increases,
as found in experiment,\cite{frolov2} and occur predominantly in regions
where $T_c$ is low. Using the 
$T_c$ calculated from the linearized theory and the $0\leftrightarrow\pi$ phase transitions,
we obtain the full phase
diagram in an extended region of parameter space 
spanned by
$T$, $d_F$, and $\Lambda$.
We compare our calculated oscillations in $T_c$ as a function of $d_F$
with reported Nb/Co experimental data\cite{obi99} and find good agreement.

\section{\label{sec:methods}Methods}

The  systems that we study
consist of slabs of clean superconductor (S) material separated by 
ferromagnetic  (F) layers.
We will emphasize  trilayers consisting of one SFS junction
and, as a sample of what can generally occur in multilayers, present
also results for seven layer systems consisting of three SFS junctions.
The thickness of the S layers in an SFS junction is denoted by $d_S$,
and that of the F layers  by $d_F$. The  seven layer
system consists of three SFS junctions stacked 
together, so that the thickness of the two inner S layers is
$2d_S$. 
We  assume that the layers are
semi-infinite in the directions perpendicular to the interfaces
(the $x-y$ directions) and that the interfaces
are sharp. The spatial
inhomogeneity is confined to the $z$ direction, allowing us to model the 
system as quasi one dimensional. 
We assume parabolic bands, thus in the transverse direction
$\epsilon_\perp={k_\perp^2}/{2m}$, where $k_\perp$ is the wavevector in the
transverse direction and $\epsilon_\perp$ is the energy corresponding to the
$x-y$ variables.

We use the microscopic Bogoliubov-deGennes\cite{dg}  equations
to study this inhomogeneous system.  Given a pair potential (order parameter)
$\Delta(z)$ that is to be determined self
consistently, the spin-up quasi-particle ($u_n^\uparrow(z)$) and spin-down 
quasi-hole ($v_n^\downarrow(z)$) amplitudes obey the BdG
equations in the following form:
\begin{widetext}
\begin{equation}
\left( \begin{array}{cc}
H-h(z) & \Delta(z) \\
\Delta(z) & -(H+h(z)) \end{array} \right)
 \left( \begin{array}{c}
u_n^\uparrow(z) \\
v_n^\downarrow(z)
\end{array} \right)
= \epsilon_n
 \left( \begin{array}{c}
u_n^\uparrow(z) \\
v_n^\downarrow(z)
\end{array} \right).
\label{eq:BdG}
\end{equation}
\end{widetext}
Here, $H={p_z^2}/{2m} - E_F(z) + \epsilon_\perp$ is a  single-particle
Hamiltonian where ${p_z^2}/{2m}+\epsilon_\perp$ is the kinetic energy term.
The  continuous variable $\epsilon_\perp$ 
is  decoupled 
from the $z$ direction but of course it
affects the eigenvalues $\epsilon_n$. 
We describe the magnetism by an exchange field $h(z)$ which takes the value
$h_0$ in the F material and vanishes in S.
Within the superconducting layers, $E_F(z)$ is equal to $E_{FS}$, the Fermi 
energy of the S layers measured from the bottom of
the band, while in the ferromagnet we have
$E_F(z)=E_{FM}$ so that in the F regions the 
up and down band widths are $E_{F\uparrow}= E_{FM}+h_0$ and
$E_{F\downarrow}= E_{FM}-h_0$ respectively. 
In the seven layer case we assume parallel orientation of
the magnetization in all F layers.
One should
not assume that $E_{FM}=E_{FS}$ and we therefore introduce the
dimensionless Fermi wavevector mismatch parameter $\Lambda$ by
$E_{FM}\equiv\Lambda E_{FS}$. Usually, one has $\Lambda<1$.
It is also convenient to introduce the dimensionless magnetic strength
variable $I$ by $h_0 \equiv E_{FM}I$. The $I=1$ limit corresponds
to the ``half-metallic'' case.  
We  neglect interfacial scattering.  
The amplitudes $u_n^\downarrow(z)$ and $v_n^\uparrow(z)$ 
can be written down from symmetry 
relations.\cite{dg} 

The required self-consistency condition for the pair potential $\Delta(z)$ is:
\begin{equation}
\Delta(z) = \frac{g(z)}{2}
{\sum_n}^\prime
\left( u_n^\uparrow(z)   v_n^\downarrow(z)  +
       u_n^\downarrow(z)   v_n^\uparrow(z) \right)
\tanh\left(\frac{\epsilon_n}{2T}\right)
\label{eq:sc}
\end{equation}
where here and below the prime indicates a summation over states for which 
$\left|\epsilon_n\right| \le \omega_D$, where
$\omega_D$ is the usual cutoff ``Debye'' energy and it is  understood 
that the  index $n$ includes $k_\perp$ as well as the
longitudinal variables.  The BCS coupling  $g(z)$ is taken to be a constant
$g$ in the superconductor and zero in the ferromagnet.

\subsection{\label{sub:self}Self-Consistent Solutions}

Equations~(\ref{eq:BdG})~and~(\ref{eq:sc}) comprise a non-linear 
set of equations.
Exact solutions to this set must be computed in a self-consistent manner.  
We follow the procedure\cite{hv02,hv04,bhv06} used in previous work; we omit
the repetition of the technical details.
We begin with an assumed form for $\Delta(z)$, either from a prior calculation
with similar parameters  
or an {\em a priori} guess (usually a stepwise function), and then numerically
solve Eq.~(\ref{eq:BdG}) for every value of $\epsilon_\perp$ in the
appropriate range to compute  
$u_n^\uparrow(z)$, $v_n^\downarrow(z)$, and $\epsilon_n$. 
An expansion of all quantities in terms of sine waves is used\cite{hv02} to
carry out the solution.  The required matrix elements are given, for
our geometries, in Ref.~\onlinecite{hv04}.
This resulting energy spectrum and quasiparticle amplitudes are used in
Eq.~(\ref{eq:sc}) to compute a new $\Delta(z)$.  We then feed
this new $\Delta(z)$ back into Eqs.~(\ref{eq:BdG})
and repeat this process
until the fractional difference between the average of successive solutions for 
$\Delta(z)$ is less than a threshold value that we take to be $10^{-5}$.
Solutions obtained in this way are exact up to the chosen numerical precision.

The self-consistent solution for a trilayer SFS junction
can be of the $\pi$ or the $0$ type, with the pair potential
either changing or not changing sign across the F layer, respectively.
More complicated situations can occur
in multilayers: for a three junction 
system one  can encounter 
four symmetric states ($000$, $0\pi0$, 
$\pi0\pi$, $\pi\pi\pi$, with each symbol corresponding to the state of each
junction). 
When, for a given temperature and set of geometrical and 
material parameters such as $I$, $d_F$,
and $\Lambda$, 
several\cite{hv02,hv04,bhv06} different 
self-consistent solutions, that is, local minima in the free energy, exist,
the stable state must be determined by comparing the condensation 
free energies of the competing
self-consistent states.  
As discussed in Sec.~\ref{sec:intro}, when the equilibrium state
changes as a function of
temperature~\cite{bhv06} a first order phase transition can occur,
with a corresponding
latent heat.  One of the chief goals of this paper is to 
study an extended region of parameter space, locating where such
transitions exist and then mapping out the
corresponding phase diagram.

To evaluate the free energy, $\cal F$, of the
self-consistent states
we use the formula from Ref.~\onlinecite{kos}: 
\begin{widetext}
\begin{equation}
{\cal F}(T) = -2T{\sum_n}^\prime
 \ln\left[2\cosh\left(\frac{\epsilon_n}{2T}\right)\right] +
\frac{1}{d}\int_0^d\frac{\Delta^2(z)}{g(z)}dz,
\label{eq:free}
\end{equation}
\end{widetext}
where $d$ is the total thickness of the system in the $z$-direction.
In this expression, only the energy eigenvalues appear explicitly, the
eigenfunctions appearing only indirectly through the 
self-consistent $\Delta(z).$
It is equivalent to several other expressions
found in the literature  which contain the
quasi-particle amplitudes explicitly, but it is computationally much 
more convenient.

The condensation free energy is defined as
$\Delta{\cal F}(T)\equiv{\cal F}_S - {\cal F}_N$, where ${\cal F}_S$ is the
free energy of the superconducting state and ${\cal F}_N$ that of 
the non-superconducting system.
We compute ${\cal F}_N$ by setting $\Delta=0$ in equations~(\ref{eq:BdG}) 
and~(\ref{eq:free}).  Calculating
$\Delta{\cal F}(T)$ is a significant numerical challenge: recall
that in a bulk superconductor\cite{tink} 
$\Delta{\cal F}(0)=-(1/2) N(0)\Delta_0^2$, where $N(0)$ is the usual density
of states and $\Delta_0$ is the order parameter for the bulk superconductor 
at $T=0$, 
which is several orders of magnitude smaller than 
${\cal F}_N \propto N(0)\omega_D^2$.  
Hence, to obtain $\Delta{\cal F}$  we must
subtract two numerically obtained large quantities
in order to extract a difference several 
orders of magnitude smaller than the terms
subtracted. Furthermore, as we shall see, the difference in condensation
free energies of competing self-consistent states (when they
occur) is only a small fraction of the condensation free energy
of each of them. To obtain sufficiently accurate values of $\Delta{\cal F}$
requires therefore
a very high degree of precision 
in calculating ${\cal F}_S$ and ${\cal F}_N$  
so that
we can distinguish the relatively small differences between competing states to
locate phase transitions.
This situation is made more challenging by the need to calculate
derivatives of $\Delta{\cal F}$ to obtain thermodynamic
functions and latent heats.

\subsection{Calculation of $T_c$: Linearized Solution}
While the transition temperature $T_c$ from the non-superconducting
to the superconducting state can be numerically calculated 
as the temperature at which $\Delta{\cal F}$ vanishes,
it is much easier to evaluate $T_c$ 
by treating $\Delta(z)$ as a small 
parameter and linearizing the equations. In this way the calculation
is nearly entirely analytic. The amplitudes are written as
$u_n^\uparrow(z) = u_n^0(z) + u_n^\prime(z)$ and
$v_n^\downarrow(z) = v_n^0(z) + v_n^\prime(z)$
(we have dropped the spin indices for simplicity).
The $u_n^0(z)$ and $v_n^0(z)$ terms are computed from the zeroth  order
equation, which is obtained by setting $\Delta(z)=0$ in Eq.~(\ref{eq:BdG}).
The form of the zeroth order equation implies that $u_n^{0}(z)$ and
$v_n^{0}(z)$ are completely decoupled and have distinct energy spectra, 
denoted by $\epsilon_n^p$ and $\epsilon_n^h$ respectively.
Proceeding to calculate the lowest order corrections, we incorporate
quasiparticle coupling through the pair potential matrix.
One can then obtain $u_n^\prime(z)$ and $v_n^\prime(z)$ from textbook
perturbation formulas. 
The intermediate sums are in principle over the entire zeroth order spectrum.  
but as a practical matter 
it is enough to include in these  sums 
energies $\epsilon_m^p$ and $\epsilon_m^h$ 
within a few 
$\omega_D$ of the Fermi level. 

We then expand the quasiparticle amplitudes and their first order corrections in a
sine wave basis
$\phi_q(z)$, e.g.  $u_n^0(z) = \sum_q^N u_{qn} \phi_q(z)$
and
  $v_n^\prime(z) = \sum_q^N v_{qn}^\prime\phi_q(z)$, where
$\phi_q(z)=\sqrt{{2}/{d}}\,\sin(k_q z)$, with $k_q={q\pi}/{d}$. 
The range of the sums over $k_q$ is formally infinite,  but again
it is only necessary to sum up
to a wavenumber $k_N$ with an associated energy 
a few $\omega_D$ from $E_F$. 
Inserting these expansions 
into Eq.~(\ref{eq:sc}) gives the lowest order correction to $\Delta(z)$, 
which we then expand in the $\phi_q(z)$ basis.
Upon taking into account the orthogonality of the basis functions, 
the expanded Eq.~(\ref{eq:sc})
is then transformed into the matrix equation
\begin{equation}
\Delta_l=\sum_kJ_{lk}\Delta_k,
\label{eigen}
\end{equation}
where the $\Delta_k$ are the expansion coefficients of $\Delta(z)$ in
terms of $\phi_k(z)$. 
One finds after straightforward algebra: 
\begin{widetext}
\begin{eqnarray}
J_{lk}=\frac{gN(0)}{4\pi}\int d\epsilon_\perp {\sum_n}^\prime
\left\{
\sum_m\sum_{pq}^N u_{pn}^0v_{qm}^0K_{pql} 
\frac{\sum_{ij}^Nv_{im}^0u_{jn}^0K_{ijk}}
{\epsilon_n^p-\epsilon_m^h}
\tanh\left(\frac{\epsilon_n^p}{2T}\right)
+ \nonumber \right. \\
\left. \sum_m\sum_{pq}^N v_{pn}^0u_{qm}^0K_{pql} 
\frac{\sum_{ij}^Nu_{im}^0v_{jn}^0K_{ijk}}
{\epsilon_n^h-\epsilon_m^p}
\tanh\left(\frac{\epsilon_n^h}{2T}\right)
\right\}.
\end{eqnarray}
\end{widetext}
Here we have used $gK_{ijk} = \int_{0}^{d} {g(z)}\phi_i(z)\phi_j(z)\phi_k(z)dz$.
The integral over $\epsilon_\perp$ 
reflects the dependence
of the zeroth order  quasiparticle amplitudes and energies  on
$\epsilon_\perp$, and the sum over $n$ is here
only over longitudinal quantum numbers with the prime
denoting the  limitation indicated below Eq.~(\ref{eq:sc}) 
on the energies 
$\epsilon_n^p$ and $\epsilon_n^p$.

The transition temperature can then
be found, in analogy with standard procedures,\cite{allen,lv} by treating Eq.~(\ref{eigen})
as an eigenvalue equation for the matrix $J_{lk}$. At the transition temperature $T_c$
the largest eigenvalue is unity, while 
if $T>T_c$ all eigenvalues are less
than unity.  
Unlike the free energy method described in
subsection~\ref{sub:self}, this procedure does not 
require an iterative process and only the last step (finding
the eigenvalue) must in practice be performed numerically.
Therefore this method  is much more efficient, and it
also provides a check on the numerics of our 
free energy.  

\subsection{\label{sub:etc}Other quantities}
From the self-consistent amplitudes and an energy
spectrum we can also calculate other quantities of interest such as the
density of states (DOS) and the magnetization.  The local density of states is
\begin{equation}
N(z,\epsilon)=-\sum_{\sigma}{\sum_n}\left[
\left[u_n^\sigma(z)\right]^2f^\prime(\epsilon-\epsilon_n)+
\left[v_n^\sigma(z)\right]^2f^\prime(\epsilon+\epsilon_n)\right],
\label{eq:dos}
\end{equation}
where $\sigma$ denotes spin and $f^\prime(\epsilon)$ is the first derivative 
of the Fermi function.  
One can also omit the sum over $\sigma$ and obtain the
spin dependent DOS.

Similarly, we have the average number density for each spin
subband,
\begin{equation}
\left<n_\sigma(z)\right>=\sum_n\left\{\left[u_n^\sigma(z)\right]^2f(\epsilon_n)
+\left[v_n^\sigma(z)\right]^2\left[1-f(\epsilon_n)\right]
\right\}.
\label{eq:spol}
\end{equation}
This leads to the dimensionless magnetization, $M(z)$,
\begin{equation}
M(z)=\frac
{\left<n_\uparrow(z)\right>-\left<n_\downarrow(z)\right>}
{\left<n_\uparrow(z)\right>+\left<n_\downarrow(z)\right>},
\label{eq:mz}
\end{equation}
which reduces to 
$M(z)=\left[(1+I)^{3/2}-(1-I)^{3/2}\right]/
\left[(1+I)^{3/2}+(1-I)^{3/2}\right]$
for bulk F material, within our assumptions.  This expression is in numerical 
agreement with our
results from Eq.~(\ref{eq:mz}) in sufficiently thick F 
layers.

\section{\label{sec:results}Results}


In this section we present and discuss
our results. 
As explained above, trilayers consisting of one SFS junction will
be emphasized
but results for seven layer  systems comprised
of three stacked SFS junctions will also be given
in order to show that  the  single junction results can 
indeed be generalized to multilayer samples.
We will first discuss the results for the thermodynamics
and the phase transitions that ensue. This will include a detailed discussion
of the phase diagram for the SFS trilayer in the most interesting
region of  the three dimensional
space spanned by $T$,  $\Lambda$ and the F layer thickness.
A discussion of the properties of the transition temperature $T_c$ as a 
function of $d_F$ and a comparison with experiment follow.
We will also discuss other quantities of interest, such as the 
pair amplitude, the density of states, and the magnetization. 


\subsection{Parameters and Units}
The results presented below will be given
in terms of convenient dimensionless
quantities. We measure all the lengths in units
of $k_{FS}^{-1}$, the Fermi wavevector in S. We fix $D_S \equiv k_{FS}d_S =100$.
We have taken the  BCS coherence length $\xi_0$ equal to $d_S$. 
For $d_S$ of order of or larger than $\xi_0$, results are
only weakly dependent  on $d_S$, hence our results are applicable to
a very wide range of values of this variable, provided $d_S$ is not too
small.  
The dimensionless
thickness $D_F \equiv k_{FS}d_F$ of the ferromagnetic layers will be varied
over the range of interest, which corresponds to relatively small values,
since at large ones the F/S proximity effects are negligible.
Similarly, we introduce the notation $Z\equiv k_{FS}z$.
The magnetic strength parameter is taken to be $I=0.2$ unless otherwise noted. 
The effects of varying $I$ are physically similar to those of varying 
$D_F$ since
the pair amplitude oscillations in F are
governed\cite{demler} by the difference 
$(k_\uparrow-k_\downarrow)d_F$ between Fermi wavevectors in the spin bands
in F.
The Fermi wavevector mismatch parameter, $\Lambda$,  to which results
are quite sensitive, is  varied over the experimentally
relevant  range $0.1 \le \Lambda \le 1$.
The temperature is measured in units of $T_c^0$, the critical temperature of a
bulk sample of the material S. We choose $\omega_D/E_{FS}=0.02$; 
the dimensionless quantities calculated are  not sensitive to this choice.
Condensation free energies will be given in units of $N(0)\Delta_0^2$,
twice the absolute value of the condensation free energy of a bulk 
superconducting sample of the same total thickness at $T=0$.  
A dimensionless measure of the latent heats will be given by dividing the
corresponding entropy discontinuities by  $C_n(T_c^0)$, the specific heat of a 
sample of the same overall thickness but consisting exclusively of
the S material in its normal metal state at $T_c^0$. 

We performed several checks of our numerical methods.  
We  verified
that the temperature at which the self-consistent condensation free energy
goes to zero is in each case the same as the transition
temperature obtained from the 
linearized solution to the BdG equations.
For a sample with $d_S\gg\xi_0$ and $d_F=0$, we
quantitatively recovered the well established results\cite{tink} for the
thermodynamics, including the second order phase transition at $T_c^0$ and the
associated specific heat discontinuity. 
Furthermore, the spatially averaged 
DOS computed numerically for this system  shows a
well defined gap at energies within $\Delta_0$ of $E_F$ and  the 
characteristic divergence at $E_F\pm\Delta_0$. 
This test is very severe since our numerical method must
necessarily be more accurate for smaller systems, where fewer variables
are required.
Thus the ability of our numerical
procedures to handle the relatively  large systems (over six superconducting
correlation lengths thick) considered here
is verified.  The low temperature limit was
extensively checked in Ref.~\onlinecite{hv04}, and it
was also previously verified\cite{hv02}
that our methods give the correct thickness dependence
of $\Delta(z)$ for a superconducting slab
as found in the literature\cite{sv}.

\subsection{Free energy}

The basic quantity that determines the phase transitions and the entire
thermodynamics is the condensation free energy, $\Delta{\cal F}(T)$. 
We will therefore begin our exposition 
by describing some of our results
for $\Delta{\cal F}(T)$ at a few  parameter values 
representative of the regions where the phase transition
behavior is richest. 

\begin{figure*}
\includegraphics[width=6in]{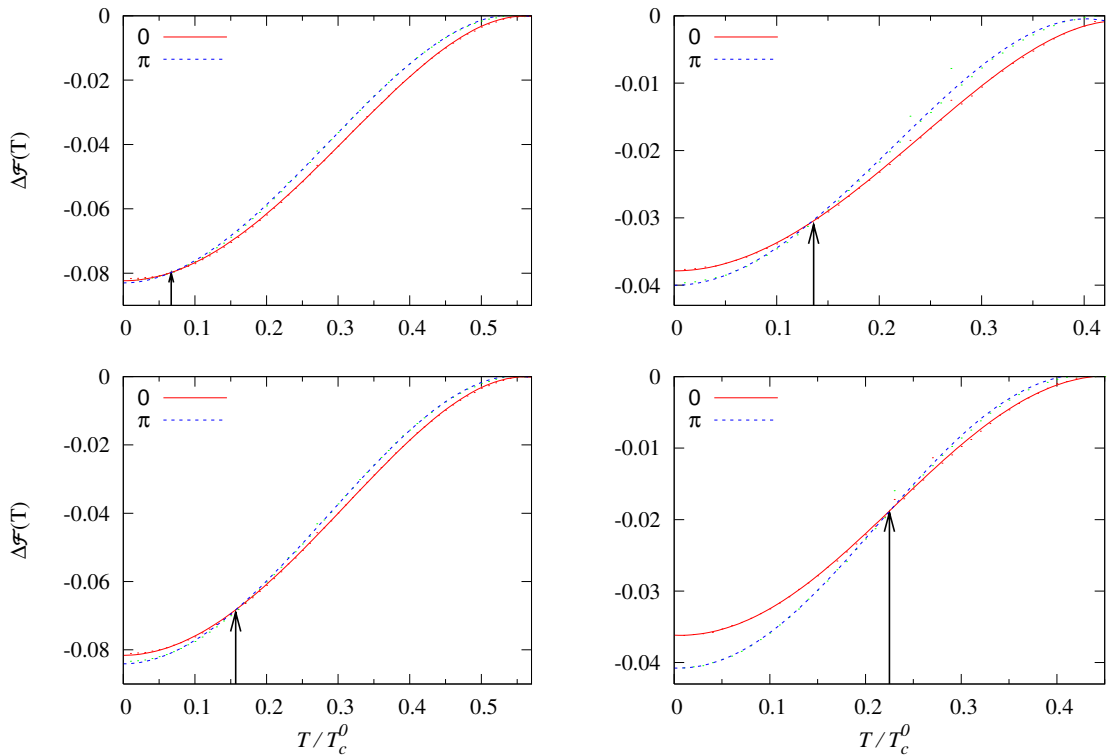}
\caption{(Color online) Results for the
normalized (see text) condensation
free energy  $\Delta{\cal F}(T)$  vs. 
temperature 
for a 3 layer SFS junction. The different curves are labeled in the legends. 
In all cases shown, upon increasing $T$ a  
$\pi$ to
$0$ transition occurs at temperature $T_x$, indicated by the arrows.
The top left panel shows results for $\Lambda=0.550$ and 
$D_F =7.0$, resulting in $T_x/T_c^0 = 0.07$.
Bottom left: $\Lambda=0.550$ and 
$D_F =7.1$, with $T_x/T_c^0 = 0.16$.
Top right: $\Lambda=0.650$ and 
$D_F =5.8$, with $T_x/T_c^0 = 0.13$.
Bottom right: $\Lambda=0.650$ and 
$D_F =5.9$, resulting in $T_x/T_c^0 = 0.23$.
}
\label{fig:ft3l} 
\end{figure*}

We begin with
Fig.~\ref{fig:ft3l}
which  
shows, for an SFS trilayer, the self-consistent 
condensation free energy $\Delta {\cal F}(T)$
plotted versus reduced
temperature $T/T_c^0$.
Data points were obtained at $T/T_c^0=0.01$ 
intervals.  
The values of $D_F$ and $\Lambda$ for which
results are shown were chosen to be such that,
at $T=0$, self-consistent solutions
of both the $0$ and the $\pi$ states are found to exist.  
In each panel, the free energies of the two competing states are
shown.  The thermodynamically stable state is 
of course the one with the lower free 
energy.  The slope of $\Delta {\cal F}(T)$ approaches zero as 
$T\rightarrow0$, which indicates that
the calculated
entropy vanishes at $T=0$ as required by the third law
of thermodynamics.  The slope also
approaches zero as $\Delta{\cal F}(T)\rightarrow0$, indicating that
the transition to the normal state is of second order, without
latent heat. 
The temperature at which this second order phase transition 
occurs, $T_c$, is the temperature at which the lower $\Delta {\cal F}(T)$ 
vanishes.  The $T_c$ found this way
agrees with the independently calculated $T_c$ using the linearized BdG
equations. The inherent finite-size and proximity effects 
cause  $T_c$ to be considerably smaller than 
$T_c^0$ in all cases.

Two outstanding features of the results shown in this figure are
the existence of a 
metastable state at all temperatures from zero up to $T_c$ and  a first
order phase transition  at an intermediate temperature: 
the free
energies of the two competing
states cross at $T=T_x$. 
In all cases shown the $\pi$
state is stable below $T_x$ and the $0$ state is stable above. The position
of $T_x$ is marked by the vertical arrow in each panel.  The value
of $T_x$ changes smoothly as $D_F$ or $\Lambda$ are changed. One can
see by comparing top and bottom panels how $T_x$ changes with $D_F$ at
constant $\Lambda$.

In Fig.~\ref{fig:ft3l} one can  see the difference in
slope between the stable state and the metastable state at $T_x$, 
particularly apparent in the right panels.   
The
existence of a metastable state and the discontinuity in the slope of the
free energy of the 
stable state (i.e., the entropy) at $T_x$ indicate the existence of a first order phase transition
with an associated latent heat. That any such transition should
be first order can also be expected from the change
of symmetry of the pair amplitude. 
For a range of parameter values including those shown  in this figure, the
phase transition behavior is exceptionally rich.
In many other regions 
of $\Lambda$ and $D_F$ parameter space the behavior
is simpler: in some  there is only one self
consistent solution to the BdG equations at $T=0$, while for other ranges of 
$\Lambda$ and $D_F$  a
metastable state is found at low temperature but it never becomes the stable
state as $T$ increases.  It is only in some regions of parameter space 
that 
$0\leftrightarrow\pi$ transitions occur as a function of $T$. 
This question will be discussed
in more detail below.

\begin{figure}
\includegraphics[width=3in]{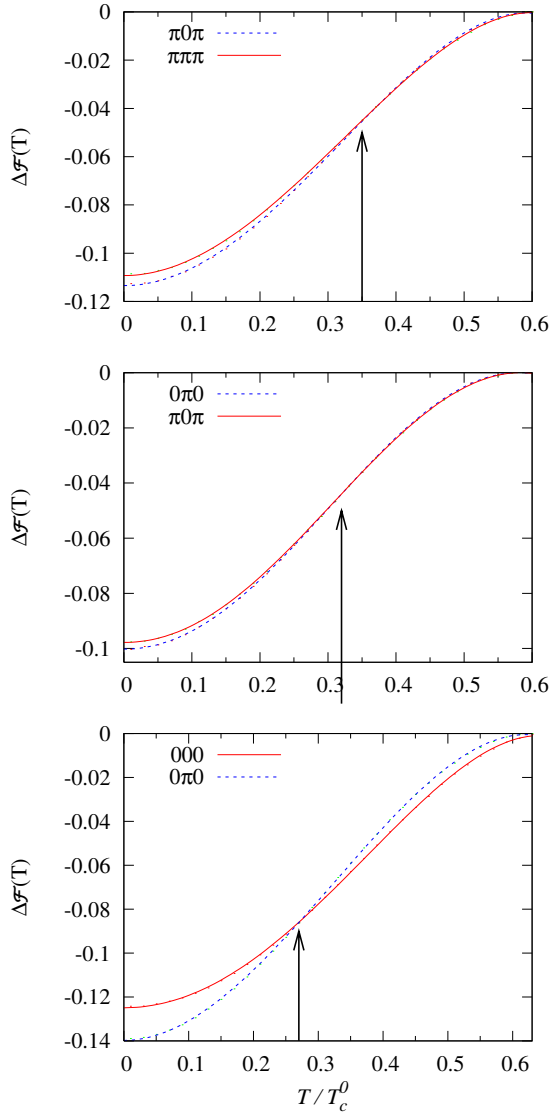}
\caption{(Color online) Results for $\Delta{\cal F}(T)$  vs. 
reduced temperature (as in Fig.~\ref{fig:ft3l}) for a 7 layer system. 
$T_x$ is indicated by the arrows.
The different curves are labeled in the legends.  
The top panel shows a $\pi0\pi\rightarrow\pi\pi\pi$ transition for $\Lambda=0.55$ and $D_F=9.1$.  The transition occurs at $T_x/T_c^0=0.37$.
The middle panel shows a $0\pi0\rightarrow\pi0\pi$ transition for $\Lambda=0.55$ and $D_F=7.9$,  at $T_x/T_c^0=0.33$.
The bottom panel shows a more pronounced $0\pi0 \rightarrow 000$ transition for $\Lambda=0.55$ and $D_F =4.75$, with $T_x/T_c^0 = 0.27$.
}
\label{fig:ft7l}
\end{figure}

Examples of similar results for the 7 layer case are shown in
Fig.~\ref{fig:ft7l}.  These  are all at  
$\Lambda=0.55$, for several values of $D_F$.  In all cases shown at least two of the four 
possible metastable states mentioned above exist over the entire temperature 
range.  The states shown  in each panel are 
the two lowest in 
free energy. In some cases additional states exist but with higher free energy 
throughout: any such states are omitted from the plots.
The three panels illustrate three different types of phase transitions
as $T$ increases:
$\pi0\pi\rightarrow\pi\pi\pi$ (one junction flipping $0\rightarrow\pi$), 
$0\pi0\rightarrow\pi0\pi$ (three flips), and
$0\pi0\rightarrow000$ (one flip, $\pi\rightarrow 0$). 
Each of these persists over a range of $D_F$.
The results show all of the same qualitative features as the three layer case:
the slope of $\Delta{\cal F}(T)$ approaches zero as  $T\rightarrow0$ and as
$\Delta{\cal F}(T)\rightarrow0$.  
There is again a change in the slope of the stable state at $T_x$
which shows that the transition is  also
of first order in multi-junction cases. 
We will show that the latent heats are of the
same magnitude as or larger than in the 3 layer case.
There is an important quantitative 
difference: $T_x$ varies more slowly with $D_F$ or
$\Lambda$ and therefore the range of parameter values for which such
transitions are found is wider.
One can expect then, that in higher order
multilayers these phenomena will be even more general.

\begin{figure}
\includegraphics[width=3in]{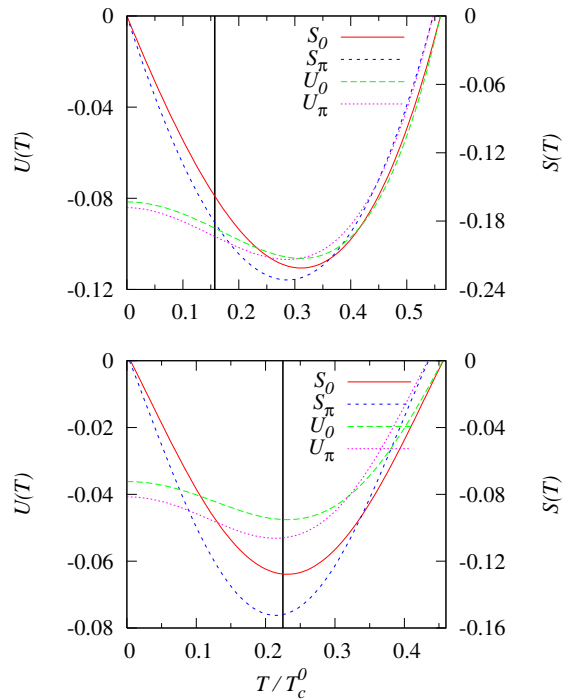}
\caption{(Color online) Thermodynamic functions of an SFS
trilayer. The top and bottom panels show the condensation energy and entropy 
(in dimensionless form, see text) for the  two sets  of  parameter
values used, respectively, in the left and right 
bottom panels of Fig.~\ref{fig:ft3l}.
The meaning of different curves is indicated
in the legend. 
The location  of the first order transition is marked
by the bold vertical line.
}
\label{fig:td3l} 
\end{figure}
\subsection{Thermodynamic functions}

From the free energy one can obtain the entire thermodynamics. 
Figure~\ref{fig:td3l} shows some of the  thermodynamic functions that 
can be obtained from  the results shown in Fig.~\ref{fig:ft3l}.
Results are shown for two quantities: the  dimensionless condensation
entropy $S(T)$, defined as  the negative derivative of 
$\Delta{\cal F}(T)$ with respect to the reduced
temperature $T/T_c^0$, and the dimensionless condensation energy $U(T)$ 
defined as $U(T)\equiv\Delta{\cal F}(T)+(T/T_c^0)S(T)$. 
Results are shown for both the stable and the metastable states as a function
of reduced temperature.    One sees that $S\rightarrow0$ smoothly 
as $T\rightarrow0$ and $T\rightarrow T_c$ in each case, which
is an important check on the computation. The condensation
entropy, energy, and free
energy all vanish at $T_c$.
In each panel a bold vertical line indicates $T_x$.  
The free energy crossings
correspond neither to crossings in $S(T)$ nor to
crossings in $U(T)$. The former follows from the phase transitions
being of first order with an associated discontinuity
in the entropy. Both the energy and
the entropy, therefore, play important roles in the phase transition.
The specific heat is not shown 
but can be calculated by taking a further derivative.

\begin{figure}
\centering
\includegraphics[width=3in]{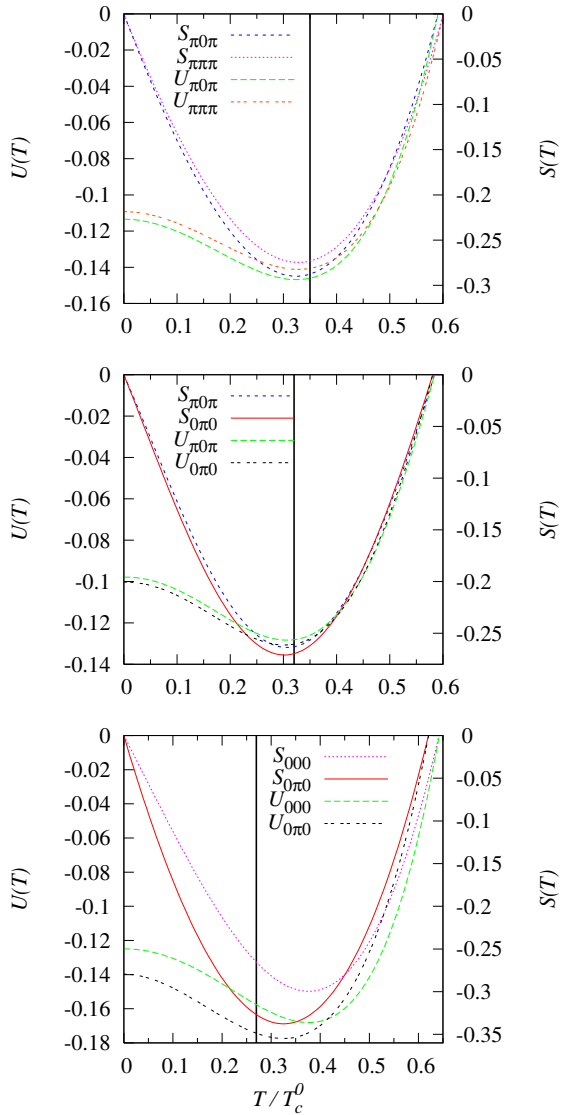}
\caption{(Color online) Thermodynamic functions of a 7 layer system. 
The panels show the condensation energy and entropy (as in
Fig.~\ref{fig:td3l}) for the parameter
sets used in Fig.~\ref{fig:ft7l}.  
The different curves are labeled
in the legend, extending the notation introduced in Fig.~\ref{fig:td3l}. 
The location  of $T_x$ is marked
by the bold vertical line. }
\label{fig:td7l}
\end{figure}

Examples of the
thermodynamic functions for the 7 layer system are shown in 
Fig.~~\ref{fig:td7l}.
The 3 junction case is again qualitatively
much like the one junction case.   The entropy, energy, and free
energy all go to zero in the appropriate limits.  There is no crossing in
energy or entropy at $T_x$, indicating the important interplay between energy 
and entropy.  The discontinuity in the entropy 
at $T_x$ also
reflects a latent heat, 
comparable to or larger than that in the one junction case.

\begin{figure}
\centering
\includegraphics[width=3in]{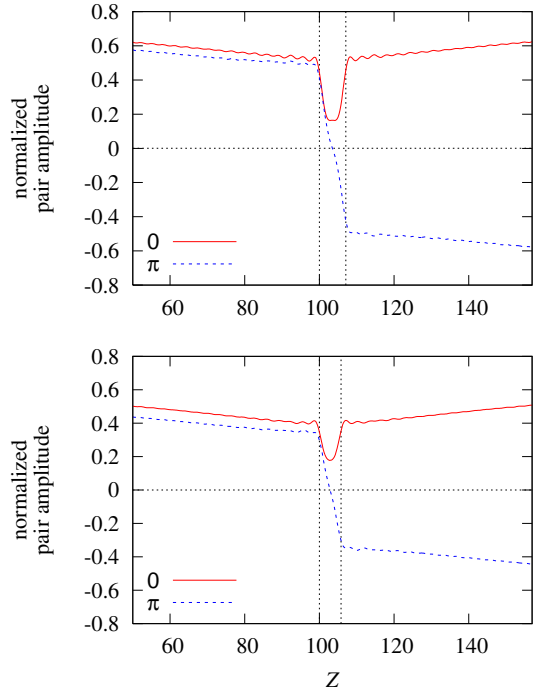}
\caption{(Color online) The normalized (see text) pair amplitude for 
the $0$ and $\pi$ states of the SFS trilayer,
at the crossing point $T_x$, as a function of position $Z\equiv k_{FS}z$.
Only the middle portion of the sample is shown. The
F layer is delimited by the vertical dotted lines.
Results are presented (top and bottom panels)
for the  two sets  of  parameter
values used, respectively, in the left and right 
bottom panels of Fig.~\ref{fig:ft3l}. 
}
\label{fig:pa3l} 
\end{figure}

The behavior of the Cooper pair amplitude
at the first order transition is illuminating. Figure~\ref{fig:pa3l}
shows, for the SFS trilayer at $T=T_x$, the pair amplitude 
(defined in the usual way as the average of spin up and down
creation operators) 
normalized to $\Delta_0/g$, its value in bulk S material at $T=0$.
Results are given versus  dimensionless position $Z$. 
The F region is in the middle, set off by vertical dotted
lines, and only  small portions of the S regions are shown.  The two 
cases shown correspond to the two bottom panels in Fig.~\ref{fig:ft3l}.
The absolute value of the pair amplitude is discontinuous at 
$T_x$: in both plots it is slightly larger for the $0$ state.
We recall that for a bulk superconductor at $T=0$, the free energy 
is proportional to the average value of the squared pair potential,\cite{tink} 
and this
is also~\cite{hv04} approximately the case at $T=0$ for 
SFS layered systems when $d_F \ll d_S \ll \xi_0$.  
Even in the bulk system, however, such a 
relationship is not valid\cite{fw} at finite temperature.
It is therefore unreasonable to expect this
proportionality in the layered case and indeed it does not occur:
the pair amplitudes do not change
continuously at $T_x$.  We conclude that the phase transitions are
not driven  by the pair amplitude. 

%
\begin{figure}
\centering
\includegraphics[width=3in]{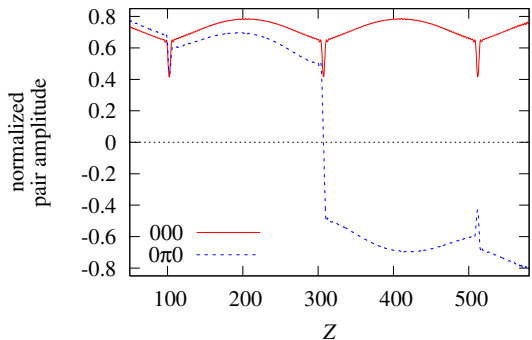}

\caption{(Color online) The dimensionless (see text) pair amplitude for the 
$0\pi0$ and $000$ states of the 7 layer system for $\Lambda=0.55$ and
$D_F=4.75$ 
at the crossing point $T_x$ as a function of position $Z$. 
This is the parameter set used in the bottom panel of Fig.~\ref{fig:ft7l}. 
}
\label{fig:pa7l} 
\end{figure}
The pair amplitude for the three junction 
system displays properties that are very similar to those of a 
single junction. 
A representative example, corresponding to
$\Lambda=0.55$ and $D_F=4.75$, is shown in Fig.~\ref{fig:pa7l}.
The absolute value of the amplitude is again discontinuous at $T_x$. 
It is very important that the 7 layer
and the 3 layer systems have qualitatively similar properties, as this
shows that the phenomena we discuss are very general.  At the same time,
in the 7 layer case there is a greater number of possible transitions and the 
regions of parameter space in which they occur
as a function of $T$ are wider, indicating
that such phenomena can be more readily found in more
complicated systems. We can make
qualitative predictions for the 7 layer system based on our quantitative
(but computationally less demanding) calculations for the 3 layer system.  Thus,
the properties of a single junction system can be generalized
to systems with many junctions.  

\subsection{Latent heats}
\begin{figure}
\centering
\includegraphics[width=3.25in]{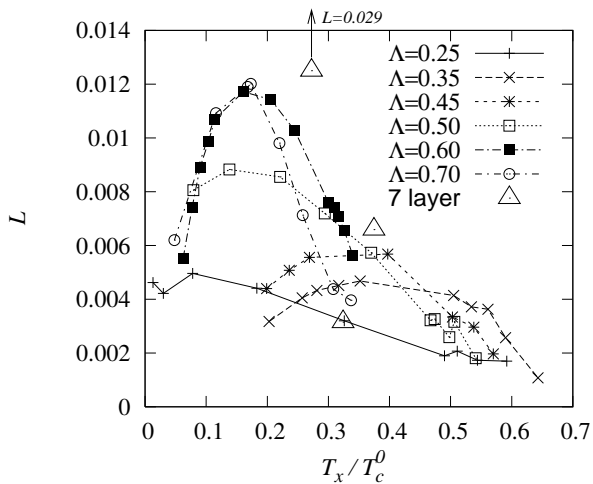}
\caption{Latent heats. 
$L$ is
the entropy discontinuity  in units of $C_n(T_c^0)$ (see text). 
It is plotted against the reduced temperature of the first 
order phase transition.  
The symbols joined by lines are  for an SFS trilayer:
the value of $T_x$ is changed 
along the horizontal axis by varying $D_F$, and from curve to curve
by varying $\Lambda$ (see legend). 
The triangles are  for the 7
layer system  cases shown in Fig.~\ref{fig:ft7l}.  
The vertical arrow attached to the topmost triangle indicates that it
corresponds to a value $L=0.029$ (off the scale).
}
\label{fig:lheat} 
\end{figure}

The signature of a first order phase transition is its
latent heat. 
In Fig.~\ref{fig:lheat} we show results for the dimensionless
latent heat $L$,
defined as   
the
difference between the entropy of the stable states
just above and  just below $T_x$ divided by
$C_n(T_c^0)$, the specific heat of a normal bulk sample of S
material  at 
$T=T_c^0$. This  is appropriate because  
$C_n(T)$ is  equal to the entropy in the free electron model. 
Results are plotted as a function of $T_x$.
Most of the results shown are for a single junction: in that case
the crossing temperature is varied by changing $D_F$
for several different values of $\Lambda$, as indicated by the
symbols connected by straight segments.
The three data points indicated by the isolated triangles correspond
to the three transitions shown in Fig.~\ref{fig:ft7l} for the 7 layer system.
One of them corresponds to a value larger, by over a factor of two, 
than the upper end of the scale.

The latent heats  
vanish as $T_x$ approaches $0$ or $T_c$,  consistent with the
smaller condensation entropy of each state in those limits. However, whenever
$T_x$  does not approach these limits the
latent heat can exceed $1\%$ of $C_n(T_c^0)$
for one junction,
and even more for the three junction system.   
Since we give $L$ in units of $C_n$, which
is an extensive property, it should be easier to observe these latent heats in
larger systems. 
A value of $L\approx 0.01$ would correspond to
picojoules in actual samples of relatively small size.~\cite{ryazanov} Such
latent heats can be readily observed via standard techniques used 
to measure specific and latent heats in 
films.\cite{cc} Even smaller specific heats can be measured using
multiple samples: attojoule level results have been reported\cite{bou}  in
electronic systems. 
We see therefore that whenever a first order transition occurs, the 
associated latent heat 
is observable.

\begin{figure*}
\centering
\includegraphics[width=6in]{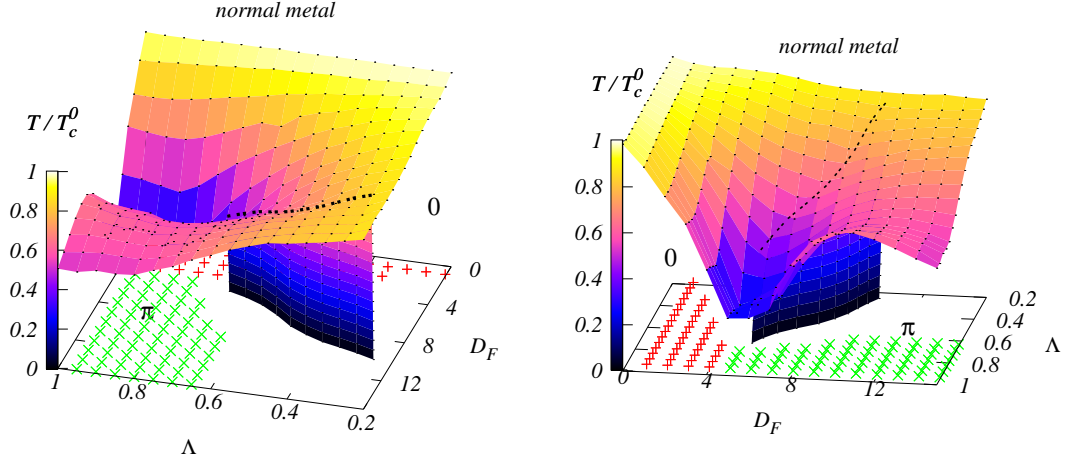}
\caption{(Color online) The ($\Lambda,D_F,T$) phase 
diagram for the 3 layer system.
The two panels show different views of the same plot.
There are three regions: in those 
labeled $0$  and $\pi$, the $0$ and $\pi$ states
are, respectively, the equilibrium state, 
while
{\em normal metal} indicates where the sample is nonsuperconducting. 
The top surface separates non-superconducting and superconducting
regions. The fairly vertical sheet marks the
temperature transitions between $0$ and $\pi$ states.
The intersection of the $0-\pi$ and the $T_c$ boundaries is
marked by a dotted line.
The portion of the $T=0$
plane marked by $\times$ symbols is the range of ($\Lambda,D_F$) 
for which only the $\pi$ state exists for all $T$: there is no metastable state 
of the $0$ type.  Likewise, in the region marked by $+$ symbols  
only the $0$ state exists. In the portion left blank, solutions
of both kinds are possible. 
}
\label{fig:phase} 
\end{figure*}

\subsection{Phase diagram}

We have seen that in an SFS trilayer there are two kinds of phase 
transitions. First, there are  second order phase transitions from the normal state
to a superconducting state of either the $0$ or  the $\pi$ kind. There
are also,  at certain ranges of the relevant 
parameters, first order transitions between  the $0$ and $\pi$ 
superconducting
states. 
As a practical matter, observability of the latter transitions 
through thermodynamic measurements requires
an appreciable difference  in condensation energies between  the two states.
This difference is an oscillatory function of $D_F$  at constant $\Lambda$ and $I$
(see e.g. figure 3 of Ref.~\onlinecite{hv04}) 
with the oscillations
becoming damped at large $D_F$, since then, 
at any $I>0$ the proximity effects are reduced and the $0$ and $\pi$ states 
are degenerate. Hence the 
most important regions theoretically and experimentally
are at relatively small values of $D_F$. As to $\Lambda$, the entire region
$\Lambda<1$ is relevant.

Therefore, we have mapped out the entire phase diagram of an SFS trilayer
in this most relevant region of $(T,\Lambda, D_F)$ space in  
Fig.~\ref{fig:phase}. 
As explained above, varying $I$  is equivalent to varying $D_F$, 
so we use $D_F$ as
the more experimentally relevant parameter. 
We show two views of the phase diagram to aid in the
visualization of this three dimensional figure.
There are three  regions in this diagram, each representing one of 
the three possible states: $0$ state, $\pi$ state, and normal (not
superconducting) state. The
crossings $T_x$ are calculated from the free energies, and $T_c$ 
through the linearization method.

The top sheet 
shows the superconductor/normal metal 
transition.  As $D_F\rightarrow0$, $T_c/T_c^0$ approaches unity for 
all $\Lambda$.  At small $\Lambda$ the sheet also flattens, since 
then the Fermi level of the ferromagnet is  small compared 
to $E_{FS}$ so that there is little interaction between the Cooper pairs and 
the ferromagnet.  
The finite temperature $T_x$ transitions between $0$ and $\pi$ regions are
located at 
the  sheet or ``wall'' that goes from the $T=0$ plane to the 
$T_c$ sheet, separating the $0$  from the $\pi$ state regions.  
This wall is of course not completely vertical:
its  deviation from verticality is what causes
first order phase transitions as a function of $T$. On the smaller
$D_F$ end, this wall ends because one of the two states becomes
unstable: a region of parameter space is entered where only one self
consistent solution exists at any temperature. Coincident with this, 
as one can see more clearly in the right panel, $T_c$
is sharply reduced: in other words, the condensation energies
of both states rise towards zero, with one actually vanishing. 
Near this region
$T_c$ has always a sharp dip. 
As one proceeds towards the opposite end of the wall, at
larger values of $D_F$, $T_c$ increases
and the wall becomes steeper, 
until  it eventually becomes  vertical. Beyond that, 
no transition occurs as a function of $T$: the stable state is the same at
all temperatures. Beyond the portion shown,
therefore, the wall would become completely vertical and
it is not depicted because it would obscure the diagram. 
It is  sufficient to show its behavior in the $T=0$ plane.
The crossings at $T=0$ are not thermodynamic phase transitions, they merely
indicate a change in the stable state as various sample parameters are
changed.


Exploration of $T_c$ and $\Delta{\cal F}(0)$ for larger values of
$D_F$ at several values 
of $\Lambda$ indicates the existence of other $0$-$\pi$ boundaries at 
larger values of $D_F$. Thus, one could extend the phase diagram in that direction,
but as previously seen\cite{hv04} and discussed
above, these additional regions are 
qualitatively similar 
to the one shown here in detail, and quantitatively less interesting.

\begin{figure}
\centering
\includegraphics[width=3.25in]{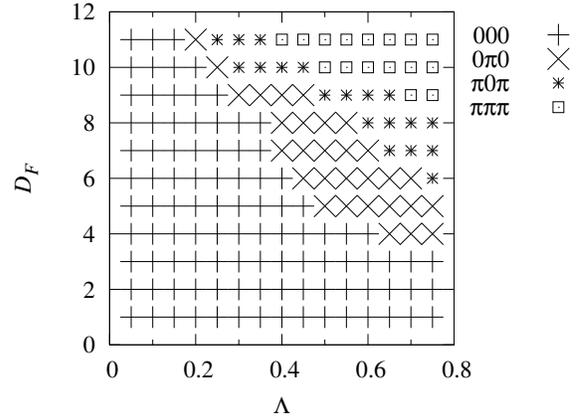}
\caption{The $T=0$ plane of the phase diagram for 
the 7 layer system.
The regions in which each of the four possible symmetric states is the
stable one are indicated by the symbols  in the legend.
There are also metastable states at most
values of $\Lambda$ and $D_F$. }
\label{fig:lowt7l} 
\end{figure}

Computing a complete three dimensional phase diagram
such as the one in Fig.~\ref{fig:phase}  for a 7 layer system would be very 
expensive in computational resources and is not necessary. 
We have already seen in connection
with Fig.~\ref{fig:ft7l} that first order transitions not only occur but are
more abundant in such systems. Further
evidence is in  Fig.~\ref{fig:lowt7l}
where we show   the zero temperature plane of the 7 layer
phase diagram. 
A different symbol marks  regions where each of the four possible
symmetric states 
is the stable one at very low $T$.  
The many boundaries between the various states and the presence
in many areas of metastable states (not marked) reflect that
there may be many first order phase transitions in the 7 layer system.  
We can 
thus infer 
that even larger structures will have rather intricate and rich  
phase diagram we found that the range of parameter space over which each type of transition
persists is much broader than in the 3 layer case.
For example, for a fixed $\Lambda=0.55$, we observed 
phase transitions (see Fig.~\ref{fig:ft7l})
between $000$ and $0\pi0$ states for $4.4 < D_F < 5.0$,
transitions between $0\pi0$ and $\pi0\pi$ for $7.75 < D_F < 8.0$ 
and transitions between $\pi0\pi$ and $\pi\pi\pi$ for $9.0 < D_F < 9.5$.
For a fixed $D_F=10$,
the $\pi0\pi\leftrightarrow\pi\pi\pi$ transition exists\cite{bhv06} for 
$0.35 \leq \Lambda \leq 0.50$.
We did not search for other transitions at
$D_F=10$.  

\begin{figure}
\centering
\includegraphics[width=3.25in]{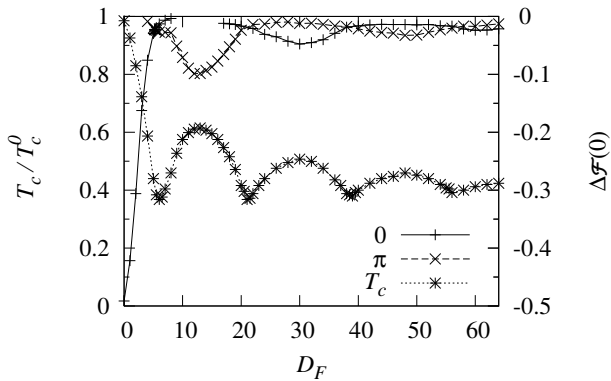}
\caption{$T_c$ vs. $D_F$  (left scale,
$*$ symbols) for $\Lambda=0.70$.
Also shown (right scale) are the energies of the $0$ and $\pi$
states at $T=0$ ($+$ and $\times$ signs respectively).  
Note the correlation between the changes in the stable
state at $T=0$ and the dips in $T_c$.}
\label{fig:tc} 
\end{figure}


By taking a slice of the  phase diagram  in Fig.~\ref{fig:phase} 
at fixed $\Lambda$, one can discern regular,
damped oscillations of $T_c$ with $D_F$.
In
Fig.~\ref{fig:tc} we show $T_c$  for $\Lambda=0.70$ over an extended
range of $D_F$.  
It is clear that as $D_F$ is increased the amplitude of the oscillations decreases. 
This
is in good agreement with experiment, as we shall see in detail below.
In addition to $T_c$, this figure shows 
$\Delta{\cal F}(0)$ 
for the $0$ and $\pi$ states. 
In a bulk superconductor, the ratio of
this dimensionless quantity to the reduced transition
temperature is  $-0.5$,
which is confirmed here by 
our result for the $0$ state at $D_F=0$. The $\pi$ state is unstable, for
obvious reasons, in the $D_F \rightarrow 0$ limit.   
At finite values of $D_F$ this relationship between normalized
condensation energy and
reduced transition temperature is 
not strictly obeyed, but there is a qualitative correlation:  
increases in the
absolute value of $\Delta{\cal F}(0)$ correspond to increases in $T_c$.
The values of $D_F$ at which the stable state switches between $0$ and $\pi$ 
correspond to the sharp dips in $T_c$ in all cases. This has also
been seen in connection with Fig.~\ref{fig:phase}
and it indicates that the structure and shape of the oscillations in $T_c$ are
strongly correlated with the low temperature state.  
The free energy data plotted have gaps, notably for the $\pi$ state near
$D_F \le 4$  and for the $0$ state at $8 \le D_F \le 17$.  
These values of $D_F$
delimit regions in which the self-consistent calculation resulted in only one 
state.  The free energy of the vanishing state goes continuously to zero
at those boundaries.  
The pair amplitude is found to go also smoothly to zero.

The low temperature crossings at the many different values of $D_F$ suggest the
location of more $0\leftrightarrow\pi$ phase transitions.
This is in agreement with the direct observations reported in 
Ref.~\onlinecite{shelukhin}.  
Another corroboration of this claim comes from Ref.~\onlinecite{frolov2}, 
in which the related parameter which they denote by $I_c$ (the 
overall critical
current of their nonuniform thickness junction)
is  found to have a 
significant dip at the $0\leftrightarrow \pi$
transition temperature.  Remarkably, these transitions were
observed in Ref.~\onlinecite{frolov2} even though
their  samples  did not have
layers of uniform thickness. 
These two experiments and others show that this is an observable and robust
phenomenon.

A similar
analysis for the 7 layer case showed the same correspondence between
$\Delta{\cal F}(0)$ and $T_c$.  However, the larger number of energy crossings at $T=0$ lead
to a closer spacing of energy crossings in $D_F$, causing several local minima in 
$\Delta{\cal F}(0)$ to appear as a single broad minimum.
The result was that multiple dips in $T_c$ often merged.  In larger systems,
the existence of a broad local minimum in $T_c$ may correspond to multiple 
crossings at low temperature.

\subsection {Comparison with experiment}
 
\begin{figure}
\centering
\includegraphics[width=3.25in]{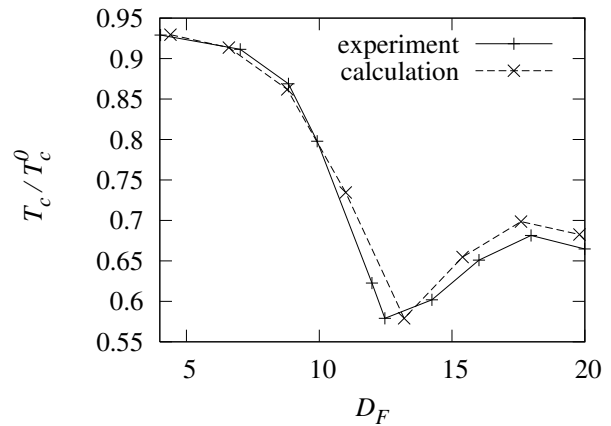}
\caption{Calculated values of $T_c$, in good agreement with 
experimental\cite{obi99} data for a Nb/Co system.
The experimental points shown are  the average of the two series 
reported in Ref.~\onlinecite{obi99}.
}
\label{fig:exp}
\end{figure}

Many experimental groups~\cite{jiang95,muhge96,jiang96,obi99,shelukhin} have
found oscillations in $T_c$ as a function of F layer thickness. Our 
calculation
also finds these oscillations 
(see Fig.~\ref{fig:tc}). The agreement with experiment is furthermore
quantitative. 
We show in
Fig.~\ref{fig:exp} a direct comparison of our results to the experimental
data for a Nb/Co system.\cite{obi99} 
In the experiment, the spontaneous magnetization of the Co layer was
found to depend on its thickness.
This means, in our language, that
$I$ must be taken as  a function of $D_F$
for the purposes of this comparison.  
To do this, we fitted 
the spontaneous magnetization reported in Ref.~\onlinecite{obi99}
and extracted, at each thickness, the value of $I$ from the formula
below Eq.~(\ref{eq:mz}).
We took a constant $\Lambda=0.60$, which
is appropriate to the materials mentioned.
The experimental and theoretical values are in excellent quantitative 
agreement on
the vertical scale, and the damped oscillations are
are well aligned in the thickness.

Comparing figures \ref{fig:exp} and \ref{fig:tc}, we conclude that
the dips in Fig.~\ref{fig:exp} must correspond to changes in the stable 
state
at zero temperature.  As these changes are, as we have seen,
associated with the first
order phase transitions, these dips in $T_c$ may also be
associated with first order phase transitions, in good agreement with
what was reported in
Ref.~\onlinecite{shelukhin}.  This implies that studies
of $T_c$ may be a useful tool for experimental discovery of first order
phase transitions and that samples which show dips
in $T_c$ are the ones that should be cooled down and studied to
locate such phase transitions.

\subsection {DOS and ${\bm {M(z)}}$}

\begin{figure}
\centering
\includegraphics[width=3.25in]{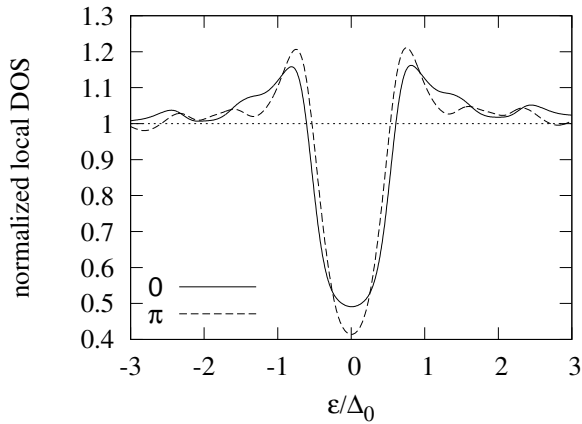}
\caption{ Density of states  at $T_x$ for an SFS trilayer. The
quantity plotted is the local DOS as defined
in Eq.~(\ref{eq:dos}),  averaged over an S layer,  
and normalized to the normal state bulk result in S.
This is  the case shown in the bottom left panel of Fig.~\ref{fig:ft3l}
($T_x/T_c^0=0.16$).
}
\label{fig:dos} 
\end{figure}



Advanced tunneling spectroscopy techniques are a useful
experimental tool to measure the local DOS, thus
probing the single-particle spectrum.
It has been found~\cite{hv04} previously that the local DOS results for $0$
and $\pi$ states are different, including a modified
subgap structure.
In such cases, tunneling
spectroscopy could be used to distinguish the states.  We now investigate
whether  the density 
of states is also a suitable technique in locating phase 
transitions.

In Fig.~\ref{fig:dos}
we show the DOS, defined as the normalized local DOS
(from Eq.~\ref{eq:dos}) averaged over one of the S layers, for
a
typical 3 layer system at
the temperature  where the first order
transition occurs.  
The case shown is for the same parameters as in the
bottom left panel of Fig.~\ref{fig:ft3l},
with $T_x/T_c^0= 0.16$.
The energy 
is normalized to the bulk S gap at zero
temperature, $\Delta_0$, while the DOS is normalized to  its
value in a bulk sample of S material
in its normal state.  
For both
$0$ 
and $\pi$ states maxima exist 
near the bulk gap edge,
qualitatively reminiscent of the divergence found 
in a bulk superconductor. The local DOS never quite goes to zero in either state, demonstrating
gapless superconductivity 
induced by the numerous
Andreev bound states in the gap. 
The
number of states in
the gap is clearly larger  for the $0$ state and the peak 
is
markedly lower. 
Although 
the DOS for both states have some general
similarities, 
the differences that do exist 
are well within
the resolution of current tunneling probes,\cite{kontos}
making
the DOS
a potentially useful
experimental technique 
in locating the phase transitions or identifying the stable state
in the neighborhood of a transition.

\begin{figure}
\centering
\includegraphics[width=3.25in]{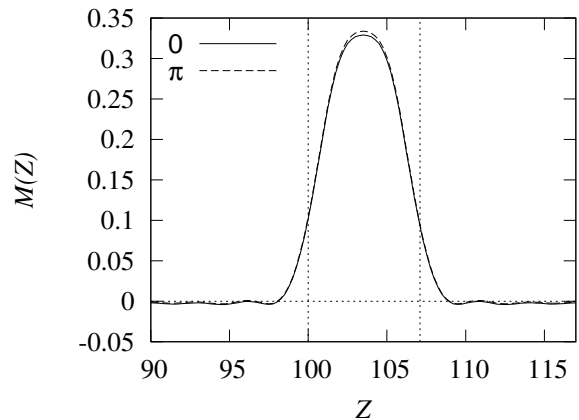}
\caption{ The dimensionless local magnetization $M(Z)$
for an SFS trilayer at $T_x$. Only
the central part of the sample is plotted. Results are
given for the two nearly coexisting states. The case
plotted is the same as in Fig.~\ref{fig:dos}. The vertical
dotted lines delimit the F region.
}
\label{fig:mz} 
\end{figure}

The last quantity we shall briefly
describe is the local dimensionless magnetization, $M(z)$, as defined
in Eq.~(\ref{eq:mz}).  Previous studies~\cite{hv04} indicate that
there is little  difference between $M(z)$ for the $0$ and $\pi$ states at
low temperature.  In that case it was found that $M(z)$ was dominated by the
exchange parameter $I$ and was
rather insensitive to the phase of the superconducting state.  
There was also little magnetization induced in the S region, as $M(z)$ decayed 
over the Fermi length scale.\cite{hv04}
To illustrate the effect that temperature has on  this trend,
we show  $M(Z)$ 
versus the dimensionless length  $Z$ at $T=T_x$
in Fig.~\ref{fig:mz}. 
In the figure, the F region is delimited by
vertical dotted lines and only a small portion of the S regions is shown.  
Consistent with Ref.~\onlinecite{hv04}, there is a quick decay and oscillation 
of $M(Z)$ in the S region.  There is a
rise in the value of $M(Z)$ to about $0.33$ in the center of the F region, 
which is
consistent with the bulk formula below Eq.~(\ref{eq:mz}) for  $I=0.2$.
Indeed, as $D_F$ increases $M(Z)$ flattens to a value
that is in good agreement with that estimate.  This is not contrary to the
experimental results in Ref.~\onlinecite{obi99} for which we modeled the change
in the saturation magnetization with $D_F$ by allowing for a $D_F$ dependent
$I$, since in that case the magnetic properties (such the
saturation magnetization) 
of the F layer were experimentally found  
to depend on $D_F$.  
Thus, the local magnetization, while interesting
for other reasons, is not a good tool for determining 
the thermodynamically stable
state or locating phase transitions.

\section{\label{sec:conclusion}Conclusion}

We have
rigorously studied the thermodynamics of clean SFS trilayer junctions
through self-consistent solutions to the BdG equations, in the clean limit.
Building on previous work\cite{bhv06} where $0\leftrightarrow\pi$ transitions
in this system were found to be possible,
we have computed  here
the three dimensional phase diagram of a clean SFS junction
over an extended and physically relevant
region of the space spanned by the parameters $T$, $d_F$,
and $\Lambda$.
We have found that
the transition to the normal state is always of second order, while
first order $\pi\rightarrow 0$  transitions  occur, as
temperature increases, over a range 
of $\Lambda$
and $D_F$. Such transitions
have been found experimentally. For systems consisting of three such junctions, we have found
here that a variety of first order transitions, involving 
$0\leftrightarrow\pi$ switching of one or more junctions, occur.
The phase transitions were shown to be driven
by a delicate balance between the condensation energy and the entropy.
The absolute value of the pair amplitude is
discontinuous at the first order
transition.
Key elements of
our approach are an efficient method to accurately compute free energies
and a 
linearization scheme that calculates $T_c$.
We have shown that dips in $T_c$ overlap with regions in parameter
space where phase transitions exist, which suggest that $T_c$ studies should be
useful for experimentally locating first order phase transitions.
We have also calculated the variation of $T_c$ with $d_F$ and found good
quantitative agreement
with an experimental study\cite{obi99} of a Nb/Co system.
We have demonstrated
that the phase transitions will have measurable latent heats, even for
relatively small samples, over a broad range of magnet thicknesses.
Another experimentally relevant quantity, the DOS,
was calculated and deemed a potentially useful tool in locating phase
transitions.
The local magnetization, however, shows little
difference between two states at the first order transition.
The method and results demonstrated here
are  expected  to be applicable
to even larger structures.

-

\section{Acknowledgments}
This work was supported in part by the University of Minnesota Graduate School
and by the ARSC at the University of Alaska Fairbanks (part of the DoD HPCM
program).


\begin{thebibliography}{0}


\bibitem{igor} I. \u{Z}uti\'{c}, J. Fabian, and S. Das Sarma, \rmp {\bf 76}, 323 (2004).
\bibitem{andreev} A.F. Andreev, Sov. Phys. JETP {\bf 19}, 1228 (1964) [Z. Eksp.  Teor. Fiz. {\bf 46}, 1823 (1964)].
\bibitem{buzdin_rmp} A.I. Buzdin, \rmp {\bf 77}, 935 (2005).
\bibitem{demler} E.A. Demler, G.B. Arnold, and M.R. Beasley, \prb {\bf 55},
15174 (1997).
\bibitem{koshina} V. N. Krivoruchko and E. A. Koshina, \prb {\bf 66}, 014521 (2002).
\bibitem{hv04} K. Halterman and O.T. Valls, \prb {\bf 69}, 014517 (2004).
\bibitem{hv02} K. Halterman and O.T. Valls, \prb {\bf 65}, 014509 (2002).
\bibitem{buzdin_dos} A. Buzdin, \prb {\bf 62}, 11377 (2000).
\bibitem{bula} L.N. Bulaevskii, V. Kuzii, and A. Sobyanin, Pis'ma Zh. Eksp. Teor. Fiz. {\bf 25}, 314 (1977) [Jetp Lett. {\bf 25}, 290 (1977)].
\bibitem{ryazanov} V.V. Ryazanov, V.A. Oboznov, A.Y. Rusanov, A.V. Veretennikov, A.A. Golubov, and J. Aarts, \prl {\bf 86}, 2427 (2001).
\bibitem{sellier} H. Sellier, C. Baraduc, F. Lefloch, and R. Calemezuk, \prl {\bf 92} 257005 (2004).
\bibitem{frolov} S.M. Frolov, D.J. Van Harlingen, V.A. Oboznov, V. V. Bol'ginov, and V.V. Ryazanov, \prb {\bf 70}, 144505 (2004).
\bibitem{guichard} W. Guichard, M. Aprili, O. Bourgeois, T. Kontos, J. Lesueur, and P. Gandit, \prl {\bf 90}, 167001 (2003).
\bibitem{jin} B. Jin, G. Su, and Q.R. Zheng, J. Appl. Phys., {\bf 96}, 5654 (2004).
\bibitem{milosevic} M.V. Milo\u{s}evi\'{c}, G.R. Berdiyorov, and F.M. Peeters, \prl {\bf 95}, 147004 (2005).
\bibitem{frolov2} S.M. Frolov, D.J. Van Harlingen, V. V. Bolginov, V.A. Oboznov, and V.V. Ryazanov, \prb {\bf 74}, 020503(R) (2006).
\bibitem{kontos2} T. Kontos, M. Aprili, J. Lesueur, F. Gen\^{e}t, B. Stephanidis, and R. Boursier, \prl {\bf 89}, 137007 (2002).
\bibitem{oboznov} V.A. Oboznov,  V. V. Bol'ginov, A. K. Feofanov, V. V. Ryazanov, and A. I. Buzdin, \prl {\bf 96}, 197003 (2006).
\bibitem{pokrovsky} V.L. Pokrovsky and H. Wei, \prb {\bf 69}, 104530 (2004).
\bibitem{obi99} Y. Obi, M. Ikabe, T. Kubo, and H. Fujimori, Physica C {\bf 317-318}, 149 (1999).
\bibitem{jiang95} J.S. Jiang, D. Davidovi\'{c}, D.H. Reich, and C.L. Chien, \prl {\bf 74} 314 (1995).
\bibitem{jiang96} J.S. Jiang, D. Davidovi\'{c}, D.H. Reich, and C.L. Chien, \prb {\bf 54}, 6119 (1996).
\bibitem{muhge96} Th. M\"{u}hge, N.N. Garif'yanov, Yu.V. Goryunov, 
G.G. Khaliullin, L.R. Tagirov, K. Westerholt, I.A. Garifullin, and
H. Zabel, \prl {\bf 77}, 1857 (1996).
\bibitem{shelukhin} V. Shelukhin, A. Tsukernik, M. Karpovski, Y. Blum, K.B. Efetov, A.F. Volkov, T. Champel, M. Eschrig, T. L\"{o}fwander, G. Sch\"{o}n, and A. Palevski, \prb {\bf 73}, 174506 (2006).
\bibitem{radovic} Z. Radovi\'{c}, N. Lazarides, and N. Flytzanis, \prb {\bf 68}, 014501 (2003).
\bibitem{radovic0} A. Radovi\'{c}, M. Lefvij, L. Dobrosavljevi\'{c}-Gruji\'{c},
 A. I. Buzdin, and J. R. Clem, \prb {\bf 44}, 759 (1991).
\bibitem{krun} B. Krunavakarn and S. Yoksan, Physica C {\bf 440}, 25 (2006).
\bibitem{buzdin} A.I. Buzdin, L.N. Bulaevskii, and S.V. Panyukov,
Pis'ma Zh. Eksp. Teor. Fiz. {\bf 35}, 147-148 (1982).
[JETP Lett {\bf 35}, 178-180 (1982)].
\bibitem{buzdin2} A.I. Buzdin and M.Yu. Kuprianov,
Pis'ma Zh. Eksp. Teor. Fiz. {\bf 53}, 308-312 (1991) [JETP Lett. {\bf 53}, 321-326 (1991)]
\bibitem{buzdin3}   A. Buzdin and I. Baladi\'{e} , \prb {\bf 67}, 184519 (2003).
\bibitem{kontos} T. Kontos, M. Aprili, J. Lesuer, and X. Grison, \prl {\bf 86}, 304 (2001).
\bibitem{beasley} S. Reymond, P. SanGiorgio, M.R. Beasley, J. Kim, T. Kim, and K. Char, \prb {\bf 73}, 054505 (2006).
\bibitem{courtois} L. Cr\'{e}tinon,  A.K. Gupta, H. Sellier, F. Lefloch, M. Faur\'{e}, A. Buzdin, and H.Courtois, \prb {\bf 72}, 024511 (2005).
\bibitem{hv03} K. Halterman and O.T. Valls, \prb {\bf 70}, 104516 (2004).
\bibitem{bhv06} P.H. Barsic, O.T. Valls and K. Halterman, \prb {\bf 73}, 144514 (2006).
\bibitem{tollis} S. Tollis, \prb {\bf 69}, 104532 (2004).
\bibitem{hv02a} K. Halterman and O.T. Valls, \prb {\bf 66}, 224516 (2002).
\bibitem{cayssol} J. Cayssol and G. Montambaux, \prb {\bf 71}, 012507 (2005).
\bibitem{buzdin4} A. Buzdin, \prb {\bf 72}, 100501(R) (2005).
\bibitem{dg} P.G. de Gennes, {\it Superconductivity of Metals and Alloys}
(Addison-Wesley, Reading, MA, 1989).
\bibitem{kos}I. Kosztin, \u{S}. Kos, M. Stone, and A.J. Leggett, \prb {\bf 58}, 9365 (1998).
\bibitem{tink} M. Tinkham, {\it Introduction to Superconductivity, 2d.
ed.} (Dover Publications, June 2004).
\bibitem{allen} P.B. Allen and R.C. Dynes, \prb {\bf 12}, 905 (1975).
\bibitem{lv} K. Levin and O.T. Valls, \prb {\bf 17}, 191 (1978).
\bibitem{sv} B.P. Stojkovi\'{c} and O.T. Valls \prb {\bf 49}, 3413 (1994);
\prb {\bf 50}, 3374 (1994), and
references therein.
\bibitem{fw} See for example, page 451 of A.L. Fetter and J.D. Walecka,
{\it Quantum Theory of Many-Particle systems}, McGraw-Hill, New York (1971).
\bibitem{cc} C.C. Huang, private communication.
\bibitem{bou} O. Bourgeois, S.E. Skipetrov, F.Ong, and J. Chaussy, \prl {\bf 94}, 057007 (2005).


\end{thebibliography}
\end{document}